\documentclass[twocolumn,twoside,reprint,superscriptaddress,aps,floatfix]{revtex4-1}
\usepackage{physics}
\usepackage{verbatim}
\usepackage{graphicx}
\usepackage[usenames]{color}
\usepackage{microtype}
\usepackage{amsmath}
\usepackage{amssymb}
\usepackage{xspace}
\usepackage{footnote}
\usepackage{relsize}
\usepackage{mhchem}
\usepackage{gensymb}

\newcommand{\bc}[1]{{#1 }}

\definecolor{brgreen}{rgb}{0,0.26,0.15}

\newcommand{\avg}[1]{\ensuremath{\left<#1\right>}}

\newcommand{\tm}{\ensuremath{T_{\text{m}}}} %
\newcommand{\kb}{k_{\rm B}} %

\newcommand{\micih}{\ensuremath{\Delta \mu^{\text{Ih}\rightarrow\text{Ic}}}} %
\newcommand{\micihcl}{\ensuremath{\Delta \mu^{\text{Ih}\rightarrow\text{Ic}}_{\text{cl}}}} %
\newcommand{\micihnn}{\ensuremath{\Delta \mu^{\text{Ih}\rightarrow\text{Ic}}_{\text{NN}}}} %
\newcommand{\micihclnn}{\ensuremath{\Delta \mu^{\text{Ih}\rightarrow\text{Ic}}_{\text{cl,NN}}}} %
\newcommand{\mic}{\ensuremath{\mu^{\text{Ic}}}} %
\newcommand{\mih}{\ensuremath{\mu^{\text{Ih}}}} %

\newcommand{\mihl}{\ensuremath{\Delta \mu^{\text{L}\rightarrow\text{Ih}}}} %
\newcommand{\mihlclnn}{\ensuremath{\Delta \mu^{\text{L}\rightarrow\text{Ih}}_{\text{cl,NN}}}} %

\newcommand{\ficm}{\ensuremath{\Tilde{m}}} %

\newcommand{\sm}{SI Appendix} %

\newcommand{\udft}{\ensuremath{U}} %
\newcommand{\uml}{\ensuremath{U_{\text{ML}}}} %
\newcommand{\gdft}{\ensuremath{G}} %
\newcommand{\hml}{\ensuremath{\mathcal{H}_{\text{ML}}}} %
\newcommand*{\meV}{\, \textrm{meV}}

\newcommand*{\meVmolecule}{\, \textrm{meV}/\textrm{H}_2\textrm{O}}

\begin{document}

\title{
Ab initio thermodynamics of liquid and solid water
}%

\author{Bingqing Cheng}
\affiliation{Laboratory of Computational Science and Modeling, IMX, \'Ecole Polytechnique F\'ed\'erale de Lausanne, 1015 Lausanne, Switzerland}
\email{bingqing.cheng@epfl.ch}

\author{Edgar A. Engel}
\affiliation{Laboratory of Computational Science and Modeling, IMX, \'Ecole Polytechnique F\'ed\'erale de Lausanne, 1015 Lausanne, Switzerland}

\author{J\"org Behler}
\affiliation{Universit\"at G\"ottingen, Institut f\"ur Physikalische Chemie, Theoretische Chemie, Tammannstr. 6, 37077 G\"ottingen, Germany}%
\affiliation{International Center for Advanced Studies of Energy Conversion (ICASEC), Universit\"at G\"ottingen, 37073 G\"ottingen, Germany}%

\author{Christoph Dellago}
\affiliation{Faculty of Physics, University of Vienna, Boltzmanngasse 5, 1090 Vienna, Austria}

\author{Michele Ceriotti} 
\affiliation{Laboratory of Computational Science and Modeling, IMX, \'Ecole Polytechnique F\'ed\'erale de Lausanne, 1015 Lausanne, Switzerland}


\begin{abstract}
Thermodynamic properties of liquid water as well as hexagonal (Ih) and cubic (Ic) ice are predicted based on density functional theory at the hybrid-functional level, rigorously taking into account quantum nuclear motion, anharmonic fluctuations and proton disorder.
This is made possible by combining advanced free energy methods and state-of-the-art machine learning techniques.
The \emph{ab initio} description leads to structural properties in excellent agreement with experiments, and 
\bc{reliable estimates of the melting points of light and heavy water.}
We observe that nuclear quantum effects contribute a crucial $0.2 \meVmolecule$ to the stability of ice Ih, making it more stable than ice Ic.
Our computational approach is general and transferable,
providing a comprehensive framework for quantitative predictions of \emph{ab initio} thermodynamic properties using
machine learning potentials as an intermediate step.
\end{abstract}

\date{This manuscript was compiled on \today}

\maketitle
Liquid \bc{water and ice are ubiquitous on Earth,
and their thermodynamic properties have important consequences
in the climate system~\cite{bartels2012ice},
the ocean,
biological cells~\cite{rall1985ice},
refrigeration and transportation systems.
}
\bc{The solid phase that is stable at ambient pressure is ice Ih, whose hexagonal crystal structure is reflected in the six-fold symmetry of snowflakes.}
The cubic form, Ic, is a metastable ice phase whose relative stability with respect to ice Ih plays a central role in ice cloud formation in the Earth's atmosphere~\cite{murray2005formation,lupi2017role,kuhs2012extent}
but is extremely difficult to measure experimentally~\cite{bartels2012ice}.

\bc{Despite the simple chemical formula, H$_2$O,}
theoretical predictions of the thermodynamic properties of liquid water and ice are extremely challenging.
This is because of 
(i) the shortcomings of common water models including conventional force-fields \cite{vega2009ice} and (semi-) local DFT approaches \cite{morales2014qmc,zhang2011structural,santra2009coupledcluster}, 
(ii) proton-disorder in ice, 
and (iii) the importance of nuclear quantum effects (NQEs)~\cite{ceriotti2018NQE}. 
In particular, calculating the chemical potential difference  $\micih = \mic - \mih$ between Ic and Ih,
which characterizes the relative stability, is extremely challenging because the zero-point configurational entropies \cite{ramirez2014configurational}, proton disorder \cite{raza2011proton} and harmonic vibrational energies of ice Ih and Ic \cite{engel2015anharmonic} differ by less than one $\meVmolecule$, so that anharmonic quantum nuclear fluctuations play a decisive role. 

Water and ice have been described with varying success invoking approximations of differing severity, including simple electrostatic dipole models for the energetics of proton-ordering \cite{lekner1998energetics}, force-field based path-integral molecular dynamics (PIMD) studies \cite{ramirez2010quantum,habershon2009competing,pamuk2012anomalous,cheng2016nuclear}, first principles quasi-harmonic (QHA) \cite{ramirez2012qha,pamuk2012anomalous}, and VSCF \cite{engel2015anharmonic,engel2018firstprinciples} studies which provide an approximate upper bound for $\micih$.
These have greatly advanced our understanding of the nature of liquid water and ice, but also highlight the harsh trade-offs between the accuracy of the description of the potential energy surface governing nuclear motion and the associated cost of sampling relevant atomistic configurations. 

In this study we 
\bc{make theoretical predictions of thermodynamic properties of ice and liquid water}
at a hybrid density-functional-theory (DFT) level of theory, taking into account NQEs, proton disorder, and anharmonicity.
This is made possible by exploiting advances in machine learning (ML) techniques to avoid the prohibitively large computational expenses otherwise incurred by extensively sampling phase space using first principles methods. 
\bc{In particular,
we employ sophisticated thermodynamic integration (TI) techniques in order to accurately and rigorously
compute the chemical potential difference between ice Ic and Ih, and between ice Ih and liquid water.}

\subsection*{First-principles thermodynamics}

As the underlying electronic structure description,
we employ the hybrid revPBE0 \cite{zhang1998gga,adamo1999pbe0,goerigk2011revpb0d3} functional with a Grimme D3 dispersion correction \cite{grimme2010d3,goerigk2011thorough}, 
which has been demonstrated to accurately predict the structure, dynamics, and spectroscopy of liquid water in molecular dynamics (MD) and PIMD simulations~\cite{marsalek2017dynamics}.
revPBE0-D3 predicts
a difference in lattice energy between the most stable proton-ordered forms of ice Ic and Ih of $U^{\textrm{Ic}}-U^{\textrm{Ih}}=-0.3\meVmolecule$ 
(see \sm~for further details), which is consistent with diffusion Monte Carlo predictions of 
$U^{\textrm{Ic}}-U^{\textrm{Ih}}=-0.4 \pm 2.9 \meVmolecule$ ~\cite{raza2011proton} and 
two different random phase approximation predictions of $-0.2\meVmolecule$ and $0.7\meVmolecule$~\cite{macher2014random}.

Since thorough sampling of the phase space of water
at the revPBE0-D3 level of theory is prohibitively expensive, we sample the phase space using a surrogate ML potential energy surface (PES), $U_{\text{ML}}$. 
We then exploit the fact that the Gibbs free energy of the surrogate systems, $G_{\text{ML}}$, can be promoted to the revPBE0-D3 level of theory using free energy perturbation 
\begin{equation}
        \gdft(p,T) - G_{\text{ML}}(p,T) 
        = - \kb T \ln\avg{ \exp\left[-\frac{\udft-\uml}{\kb T}\right]}_{p,T,\hml} \, ,
    \label{eq:fep}
\end{equation}
where $\avg{\ldots}_{p,T,\hml}$ denotes the ensemble average for the system sampled at temperature $T$ and pressure $p$ using the surrogate Hamiltonian ${\cal H}_{\text{ML}}$.
Evaluation of ~\eqref{eq:fep} 
is rendered particularly affordable and robust by the high fidelity of our surrogate machine-learning PES, which substantially exceeds that obtained from empirical force fields or local DFT calculations,
which were previously used as implicit surrogates~\cite{grabowski2009ab,glensk2014breakdown}.
\bc{
\eqref{eq:fep} is the central formula of our approach: 
it not only enables accurate and
efficient free energy estimation at the \emph{ab initio} level by delegating phase-space sampling to cheap surrogate models,
but also provides a general way for benchmarking and calibrating the ML potentials.
}

\subsection*{Neural network potential for water}
We constructed a flexible and fully dissociable neural network (NN) potential for bulk liquid water and ice following the framework of Behler and Parrinello~\cite{behler2007generalized,behler2017anie,morawietz2016van} using the RuNNer code~\cite{runner},
which was trained on the basis of revPBE0-D3 energies and forces for 1,593 diverse reference structures of 64 molecules of liquid water computed using the CP2K package~\cite{lippert1999cp2k}.
Further details regarding the DFT calculations, comparison with the energies computed using VASP~\cite{kresse1996software}, the training and validation of the NN potential can be found in the \sm.
We have released this NN potential on a public repository (to be inserted on publication).

\begin{figure}[hbt]
\includegraphics[width=0.5\textwidth]{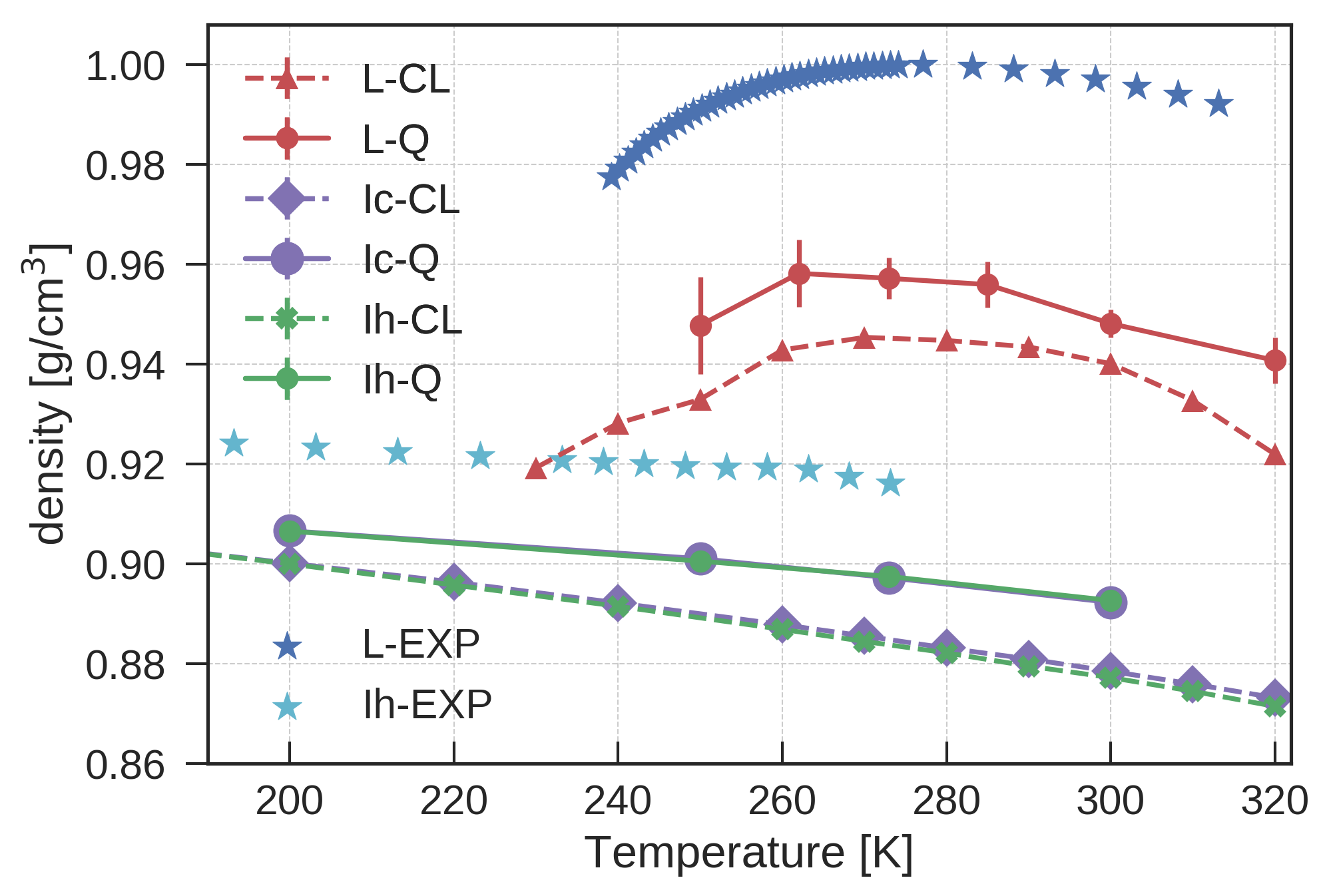}
\caption{
Classical (CL) and quantum (Q) density isobars for ice Ic, ice Ih, and liquid water (L) at $P=1$~bar computed via (PI)MD simulations using the NN potential.
The predicted densities of ice Ic and Ih almost overlap both at the quantum and the classical level.
The experimental results for undercooled water are taken from Ref.~\citenum{hare1987density}.
}
\label{fig:nn-density}
\end{figure}

\begin{figure}[hbt]
\includegraphics[width=0.5\textwidth]{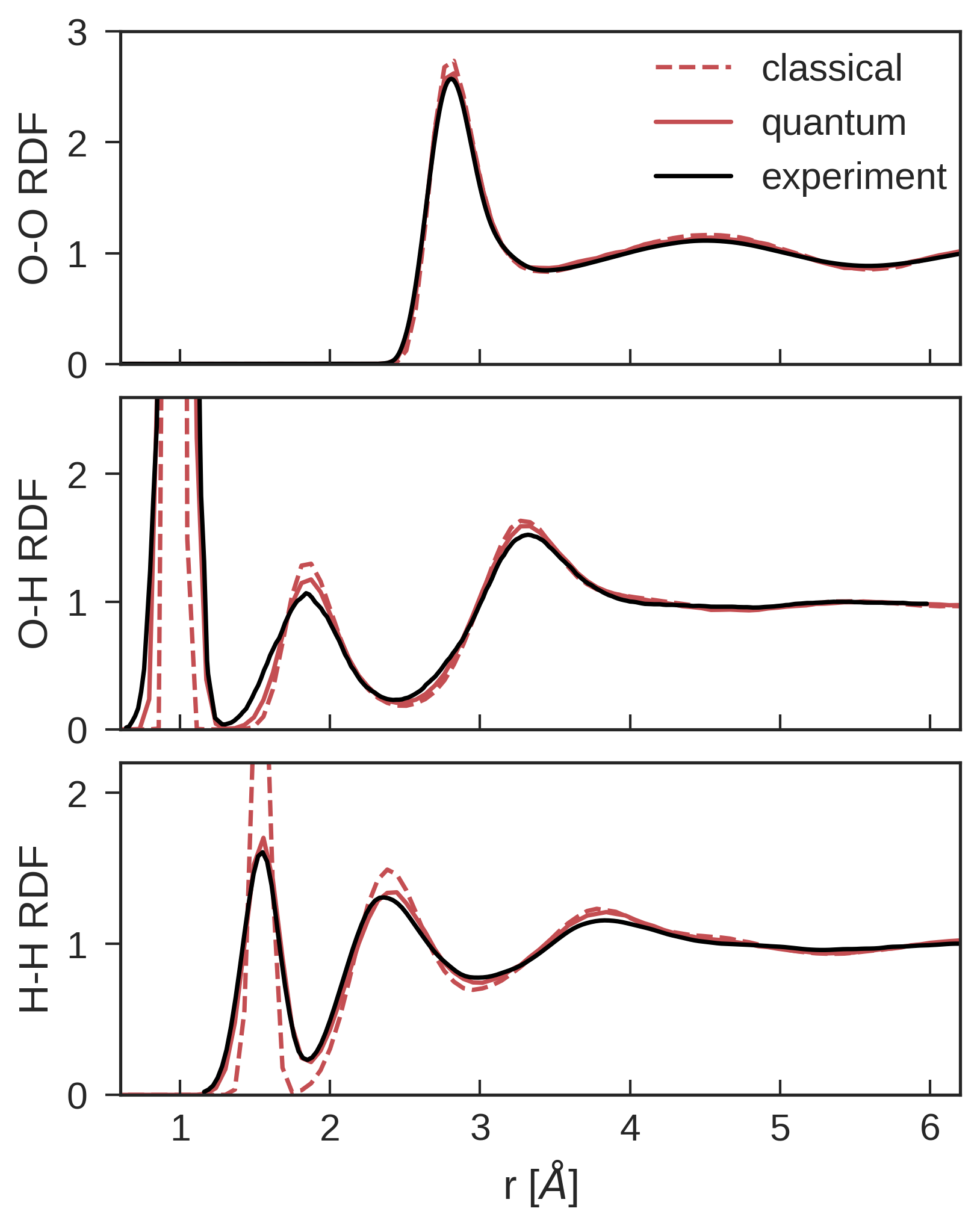}
\caption{
\bc{
Oxygen-oxygen, oxygen-hydrogen, and hydrogen-hydrogen radial distribution functions (RDF) at 300~K and experimental density computed via (PI)MD simulations in the NVT ensemble using the NN potential.
The experimental O-O RDF was obtained from Ref~\citenum{skinner2014structure},
and the experimental O-H and H-H RDFs were taken from Ref.~\citenum{soper2000radial,chen2016ab}.
}
}
\label{fig:nn-rdf}
\end{figure}

The revPBE0-D3-based NN potential
describes the density (Fig.~\ref{fig:nn-density}) and structural properties of water (Fig.~\ref{fig:nn-rdf}) in very good agreement with experiments.
Fig.~\ref{fig:nn-density} shows density isobars computed for ice Ic, ice Ih, and liquid water considering both the case of classical and quantum-mechanical nuclei.
Fig.~\ref{fig:nn-density} highlights that 
(i) the predicted densities of liquid water and ice Ih and Ic agree with experiment to within 3\%, 
(ii) the predicted thermal expansion coefficients show excellent agreement with experimental data, and
(iii) the temperature of maximum density for liquid water matches the experimental value of $3.98~\degree$C.
It also shows that NQEs lead to an increase of around 1\% in the density of the three phases of water.
This anomalous increase for the ice Ih phase has been observed in previous QHA calculations employing a number of different DFT functionals~\cite{pamuk2012anomalous}.
Experimentally, the suppression of NQEs can be partially achieved by deuteration,
and it has been observed that the molar volume of D$_2$O is 0.4\%~\cite{bridgman1935pressure} larger compared with H$_2$O for liquid water at the ambient temperature,
and about 0.3\% larger for hexagonal ice at 250\,K~\cite{roettger1994lattice}.

\bc{The upper panel of Fig.~\ref{fig:nn-rdf} shows that NQEs have a slight de-structuring effect on the oxygen-oxygen (O-O) radial distribution function (RDF),
bringing it in excellent agreement with experiment from X-ray diffraction
measurements~\cite{skinner2014structure},
as also seen in the \emph{ab initio} (PI)MD calculations with revPBE0-D3~\cite{marsalek2017dynamics}.
This de-structuring has previously been observed in simulations employing other DFT functionals~\cite{morrone2008nuclear} as well as empirical water models~\cite{mantz2006structural,ceriotti2011accelerating},
and was rationalized as a result of competing quantum effects~\cite{habershon2009competing,ceriotti2016nqe}.
The central and bottom panels of Fig.~\ref{fig:nn-rdf} further show that 
NQEs cause a significant broadening of the oxygen-hydrogen (O-H) and hydrogen-hydrogen (H-H) RDFs,
especially around their respective first peaks,
which plays a predominant role in ensuring the match between the simulations and experiment.
It is worth noting that the 
agreement between the NN and the experimental RDFs in Fig.~\ref{fig:nn-rdf} is significantly better compared to most previously benchmarked empirical water models and DFT functionals~\cite{spura2015nuclear,cisneros2016modeling}.
}

\begin{figure}[hbt]
\includegraphics[width=0.5\textwidth]{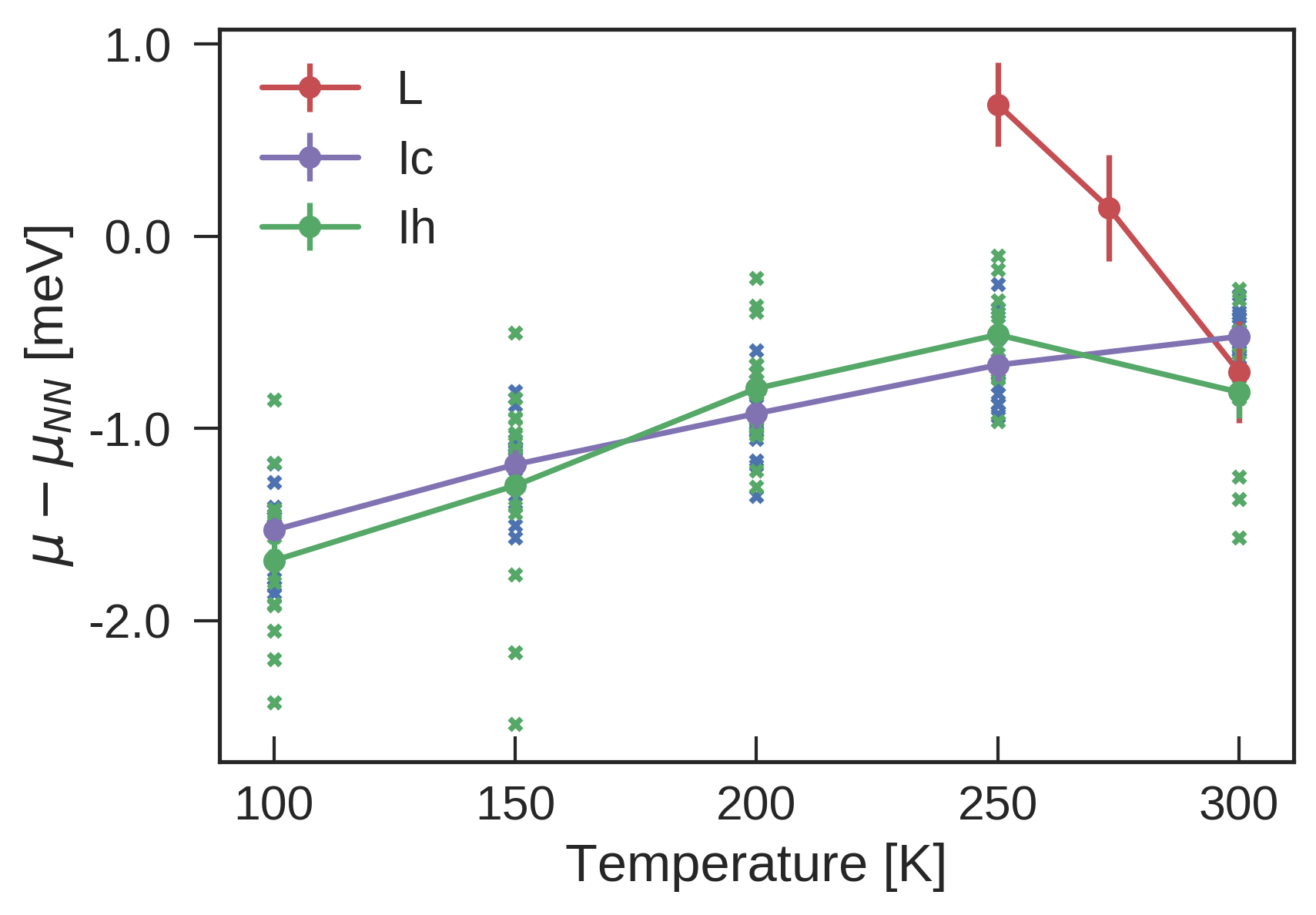}
\caption{
The difference in the chemical potential $\Delta\mu_{\textrm{NN}} \equiv \mu - \mu_{\textrm{NN}}$ between revPBE0-D3 and NN-based MD simulations at $P=1~\textrm{bar}$.
Standard errors of the mean are indicated by the error bars.
The violet (green) crosses indicate the results from 16 different 64-molecule proton-orderings of Ic (Ih).
The violet (green) line shows the average $\Delta\mu_{\textrm{NN}}$ across proton-orderings.
}
\label{fig:nn-dft-dif}
\end{figure}

\subsection*{Promoting ML potential to DFT}
\bc{Despite the excellent performance of the NN potential, }
the fitting strategy, the finite cut-off radii applied to the description of atomic environments, and possible ``holes'' in the training set~\cite{behler2015constructing} inevitably lead to small residual errors compared to the underlying first-principles reference.
To assess their significance, we have trained a collection of NN potentials using different training sets and/or initial random seeds, which demonstrates that predictions of the chemical potential difference between ice Ic and Ih from two different NN potentials can be as large as $1\meVmolecule$ (see Fig.~S4 in \sm~for further detail).
Promoting the results to the DFT level eliminates these residual errors and any dependence on the specific NN potential employed. This allows us to achieve sub-meV accuracy in free energies (as required to resolve the greater stability of ice Ih compared to Ic) and to make unbiased properties predictions at the reference \emph{ab initio} level of theory in general.

The temperature-dependent DFT corrections to the NN chemical potentials of different phases of water, $\Delta\mu_{NN} \equiv \mu-\mu_{\textrm{NN}}=(\gdft- G_{\text{NN}})/N$, as obtained from free energy perturbations (~\eqref{eq:fep}) performed on 64-molecule systems, are shown in Fig.~\ref{fig:nn-dft-dif}.
For each ice phase (Ic and Ih) 16 different proton-disordered initial configurations with zero net polarization were generated using the Hydrogen-Disordered Ice Generator~\cite{matsumoto2018genice}.
The standard deviation of the potential energy for the 16 proton-disordered ice Ic configurations is $0.3 \meVmolecule$ ($0.25 \meVmolecule$) using the NN potential (DFT), respectively.
For ice Ih it is $0.4 \meVmolecule$ ($0.25 \meVmolecule$) using the NN potential (DFT).
Starting from these different initial configurations is crucial here, 
because (i) the proton order is effectively ``frozen-in'' at the timescales available to simulation \cite{grau2014qmsimulation} and (ii)
there are significant differences between $\Delta\mu_{NN}$ of different proton-disordered states (see Fig.~\ref{fig:nn-dft-dif}).
For liquid water,
1000 single-point revPBE0-D3 calculations for un-correlated configurations generated from NN-based NPT simulations suffice to converge the value of the calibration term $\Delta\mu_{NN}^\textrm{L}$ to about $0.2 \meVmolecule$.
For each proton-disordered ice structure, 200 such single-point calculations are enough to converge $\Delta\mu_{NN}^\textrm{Ic}$ and $\Delta\mu_{NN}^\textrm{Ih}$ to $0.1 \meVmolecule$.

\section*{Results and discussions}
\subsection*{The relative stability of hexagonal and cubic ice}

\bc{We follow the workflow illustrated in Fig.~\ref{fig:ti_scheme} in the Materials and Methods section to evaluate the chemical potential difference $\micih$ at the revPBE0-D3 level of theory, taking into account  nuclear quantum fluctuations.}
\bc{We first compute the classical absolute free energies of the two ice phases at the NN level using the TI methods described in Ref.~\cite{cheng2018computing},
and thereby the corresponding chemical potential difference $\micihclnn$.}
The classical chemical potential difference between ice Ih and Ic at the revPBE0-D3 level can then be evaluated as 
$\micihcl=\micihclnn+\Delta\mu_{NN}^{Ic}-\Delta\mu_{NN}^{Ih}$.

\bc{Note that the speed and linear scaling of the NN potential allows us to simulate ice systems containing as many as 768 water molecules.
Such system size is not only essential to represent the wide spectrum of possible local arrangements realized in proton-disordered ice,
but also important for averaging over different proton disordered structures when correcting for the chemical differences between the NN potential and revPBE0-D3,
as demonstrated by the spread of $\Delta\mu_{NN}$ between different structures in Fig.~\ref{fig:nn-dft-dif}.
}

\begin{figure}[hbt]
\includegraphics[width=0.5\textwidth]{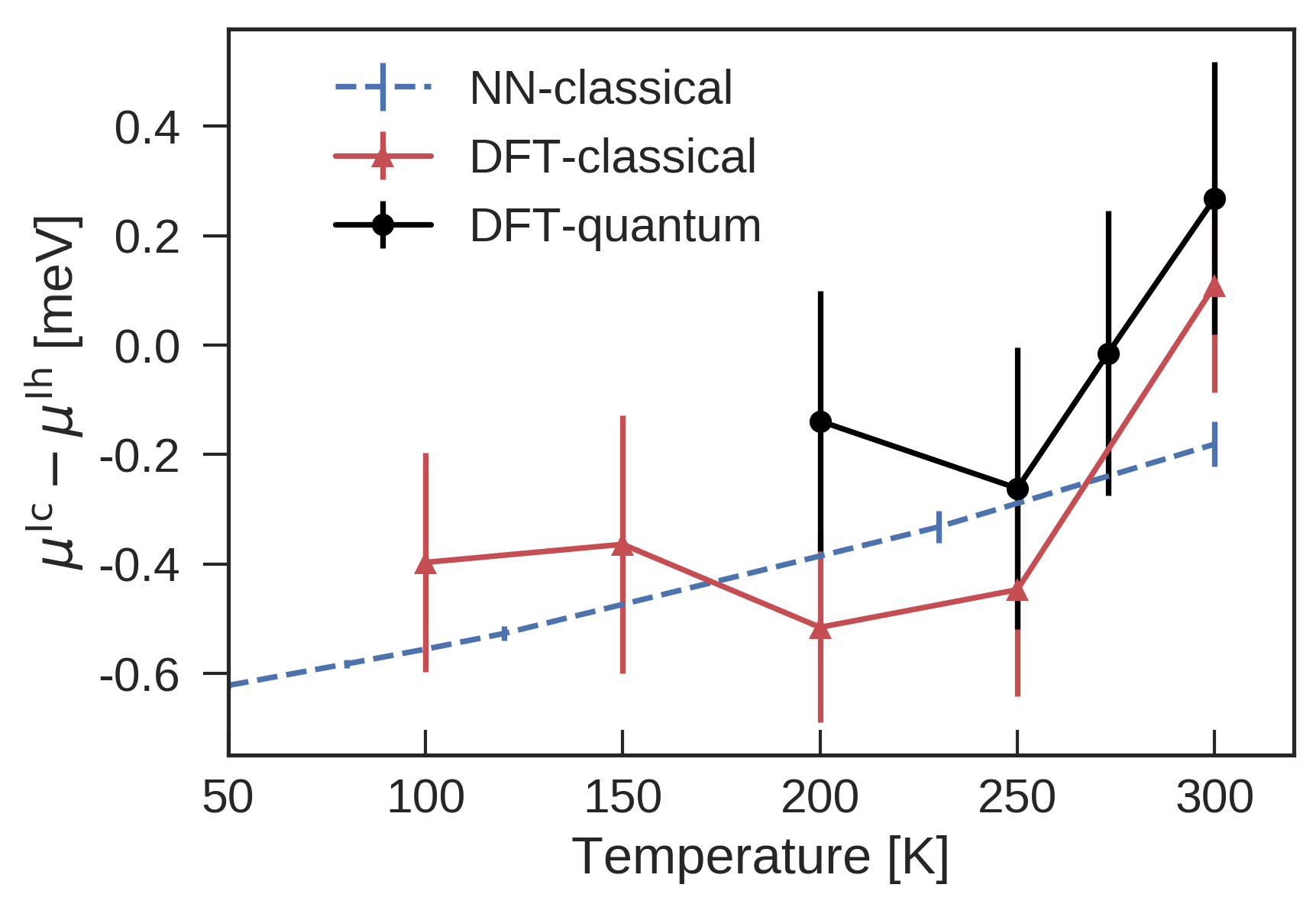}
\caption{
Temperature dependence of the chemical potential difference between ice Ih and Ic at 1 bar. The errors associated with the classical and quantum-mechanical revPBE0-D3 values arise predominantly from the differences in $\Delta\mu_{\textrm{NN}}$ between different proton-orderings.
}
\label{fig:nn-muicih}
\end{figure}

NQEs are taken into account by integrating the quantum centroid virial kinetic energy $\avg{T_{CV}}$ with respect to the fictitious  ``atomic'' mass from the classical (i.e. infinite) mass to the physical masses of oxygen and hydrogen atoms
 (see Materials and Methods and Fig.~\ref{fig:si2}).
We perform NN-based PIMD simulations within the NPT ensemble and assess the impact of NQEs on the chemical potential at the NN level using $\micihnn-\micihclnn$. 
We note that the NN potential is not ``biased'' towards Ic or Ih as the NN to revPBE0-D3 calibration terms $\Delta\mu_{NN}^\textrm{Ic}$  and $\Delta\mu_{NN}^\textrm{Ih}$ are similar (Fig.~\ref{fig:nn-dft-dif}),
and that the difference in $\avg{T_{CV}}$ of difference water phases was previously found to be very similar for three completely different inter-atomic potentials~\cite{cheng2016nuclear}.
Combining all of these terms, we finally arrive at the result $\micih=\micihcl+\micihnn-\micihclnn$.

Fig.~\ref{fig:nn-muicih} shows that the NN predictions of $\micih$ and the revPBE0-D3 results are statistically indistinguishable.
At the classical level $\micihcl$ is negative, especially at low temperatures.
Consistent with the VSCF results of Ref.~\citenum{engel2015anharmonic}, proton disorder introduces substantial variations in the chemical potential of ice Ic and Ih.
More importantly, nuclear quantum fluctuations are crucial to stabilize ice Ih. 
At the quantum-mechanical level $\micih$ is close to zero at 200-250\,K and increases to $0.2 \pm 0.2 \meVmolecule$ at 300\,K, 
suggesting ice Ih is more stable after all.
For comparison, at the classical level, the monoatomic water model~\cite{molinero2008water}  -- which omits hydrogen atoms --  predicts a negligible difference ($\micih(240~K)=0.032 \pm 0.002 \meV$~\cite{cheng2018theoretical}), while the MB-pol forcefield~\cite{reddy2016accuracy}, which includes many-body terms fitted to the coupled-clusters level of theory, predicts a small negative value ($-0.4\meVmolecule$) (see \sm~for further detail).
\bc{
Assuming that 
the heat of transition from ice Ic to ice Ih is approximately constant over the temperature range 200-300\,K,
the temperature dependence of $\micih$ implies  (using a TI with respect to $T$ analogous to ~\eqref{eq:ti-T}) a transition enthalpy of $H^\text{Ic}-H^\text{Ih}=1.0 \pm 0.5 \meVmolecule$,
consistent with
the wide experimental range $0.1-1.7\meVmolecule$~\cite{carr2014spectroscopic}.
}

\subsection*{The relative stability of hexagonal ice and liquid water}
\bc{Now we compute the difference in chemical potential $\mihl=\mu^{Ih}-\mu^{L}$ between the proton-disordered ice Ih and liquid water.
The approach is, in analogy to the schematics in Fig.~\ref{fig:ti_scheme}, to obtain the NN chemical potential difference before promoting it to the DFT level and adding NQEs.

We first compute $\mihlclnn$ using the interface pinning method~\cite{pedersen2015computing} in classical MD simulations with the NN potential.} We then fit $\mihlclnn$ from independent simulations \bc{at different temperatures} to the TI expression
\begin{equation}
    \mihlclnn(T) =- k_BT \int_{T_m}^{T} \frac{\left<H^{\mathrm{Ih}}_{\mathrm{cl,NN}}\right>_{P,T} -\left<H^{\mathrm{L}}_{\mathrm{cl,NN}} \right>_{P,T}}{k_B T^2} dT,
    \label{eq:ti-T}
\end{equation}
where $H_{cl,NN}$ is the enthalpy of the classical system described by the NN potential, whose value has been computed from separate NPT simulations (Fig.S3 in \sm).
\bc{Afterwards, the calibration terms for chemical potentials $\Delta\mu_{\mathrm{NN}}^\mathrm{L}$ and $\Delta\mu_{\mathrm{NN}}^\mathrm{Ih}$ (Fig.~\ref{fig:nn-dft-dif}) are added in order to obtain the revPBE0-D3 predictions for the classical systems.
Finally,}NQEs in H$_2$O water and D$_2$O water are considered by performing a series of PIMD simulations at different fictitious masses using the NN potential.

\begin{figure}[hbt]
\includegraphics[width=0.5\textwidth]{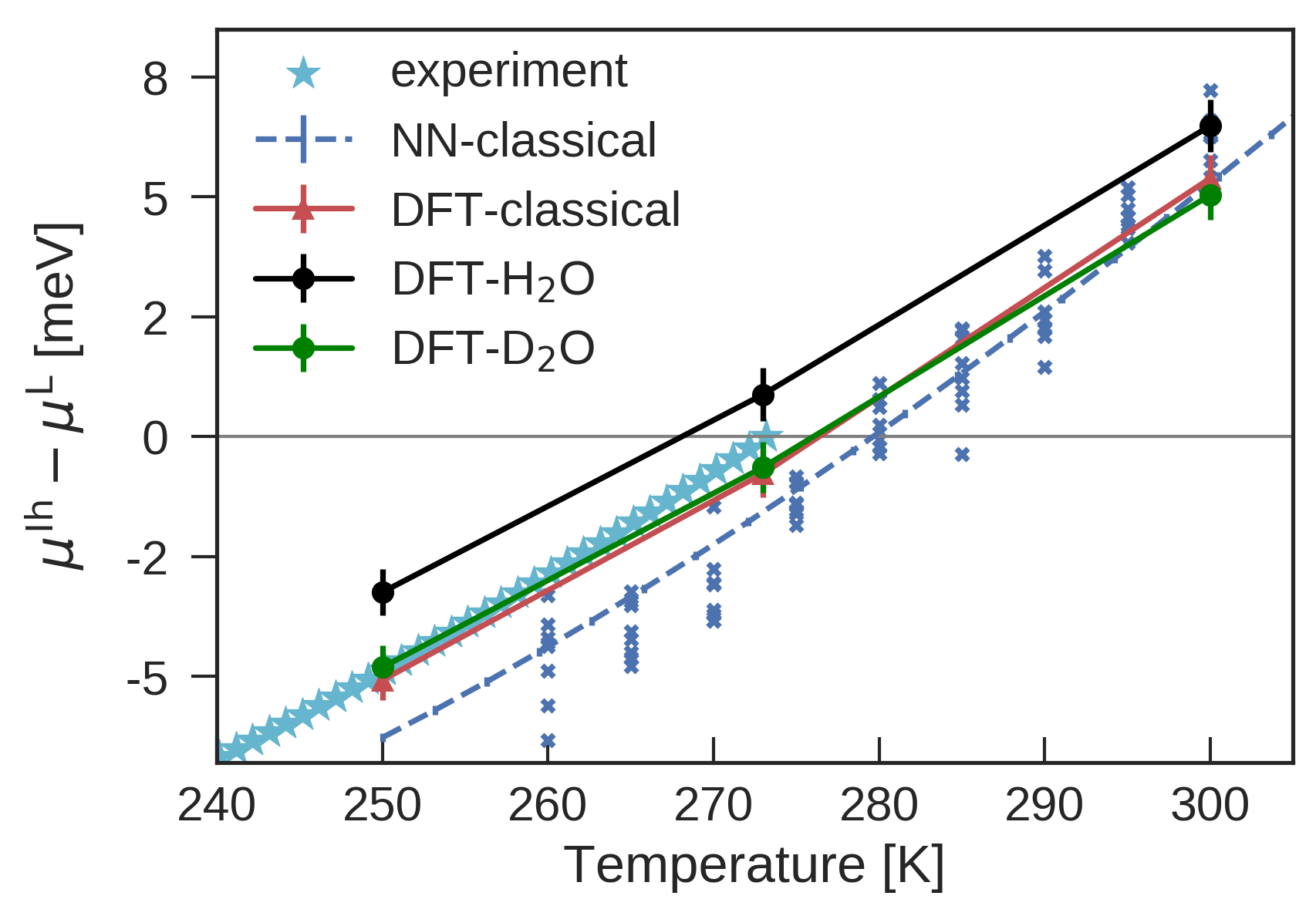}
\caption{
Temperature dependence of the chemical potential difference between liquid water and ice Ih at 1\,bar.
Blue crosses indicate $\mihlclnn$ from independent interface pinning simulations,
and the blue dashed line indicates the best fit of these results to the TI expression in Eqn~\eqref{eq:ti-T}.
\bc{The experimental values were calculated from the heat capacities reported in Ref.~\citenum{haji2015direct}.}
}
\label{fig:muih-l}
\end{figure}

\bc{
Fig.~\ref{fig:muih-l} shows $\mihl$ predicted at different levels of theory 
along with experimental data for H$_2$O~\cite{haji2015direct}.
A comparison between the melting points $\tm$ and the heat of fusion
$H_f=H^\mathrm{L}(\tm)-H^\mathrm{Ih}(\tm)$
of different models is provided in Table~\ref{tab:gm}.
For revPBE0-D3 H$_2$O water with NQEs,
the predicted $\tm$ has only about 2\% of error compared with experiment,
and the values of $\mihl$ are well within 15\% of experimental values at moderate undercoolings of $< 20~$K below $\tm$. 
$H_f$ is under-estimated using revPBE0-D3 and including NQEs,
which maybe due to the artifacts of the revPBE0-D3 functional,
or the limitations of representing proton disorders in natural ice even when using state-of-the-art methods~\cite{matsumoto2018genice}.
Overall, the predictions here constitute a substantial improvement over most commonly-used empirical water models~\cite{vega2009ice}. For instance, 
TIP4P models underestimate $H_f$ by 20-30\%
\cite{espinosa2014homogeneous}).
}

\begin{figure}
\centering
\includegraphics[width=0.45\textwidth]{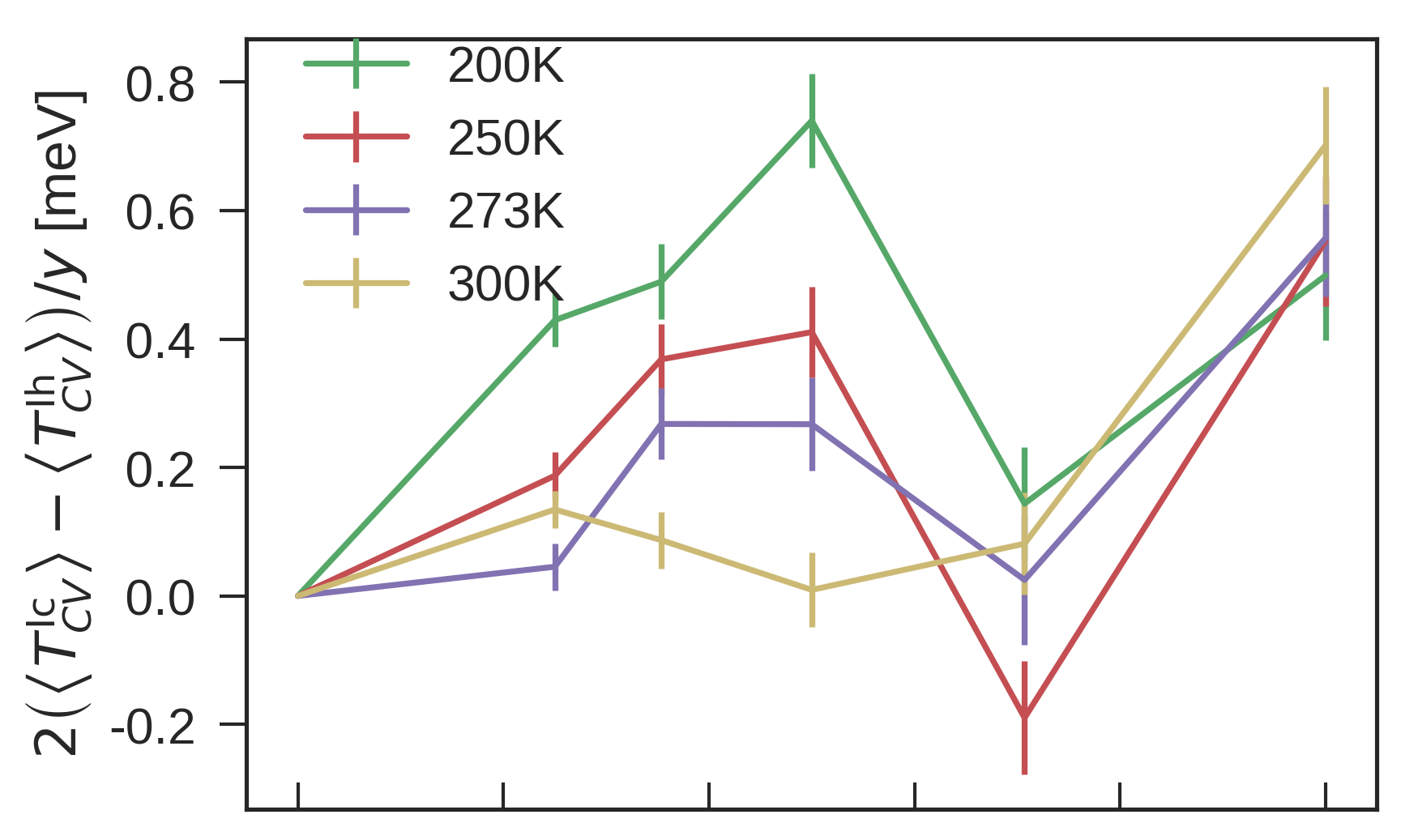}
\includegraphics[width=0.45\textwidth]{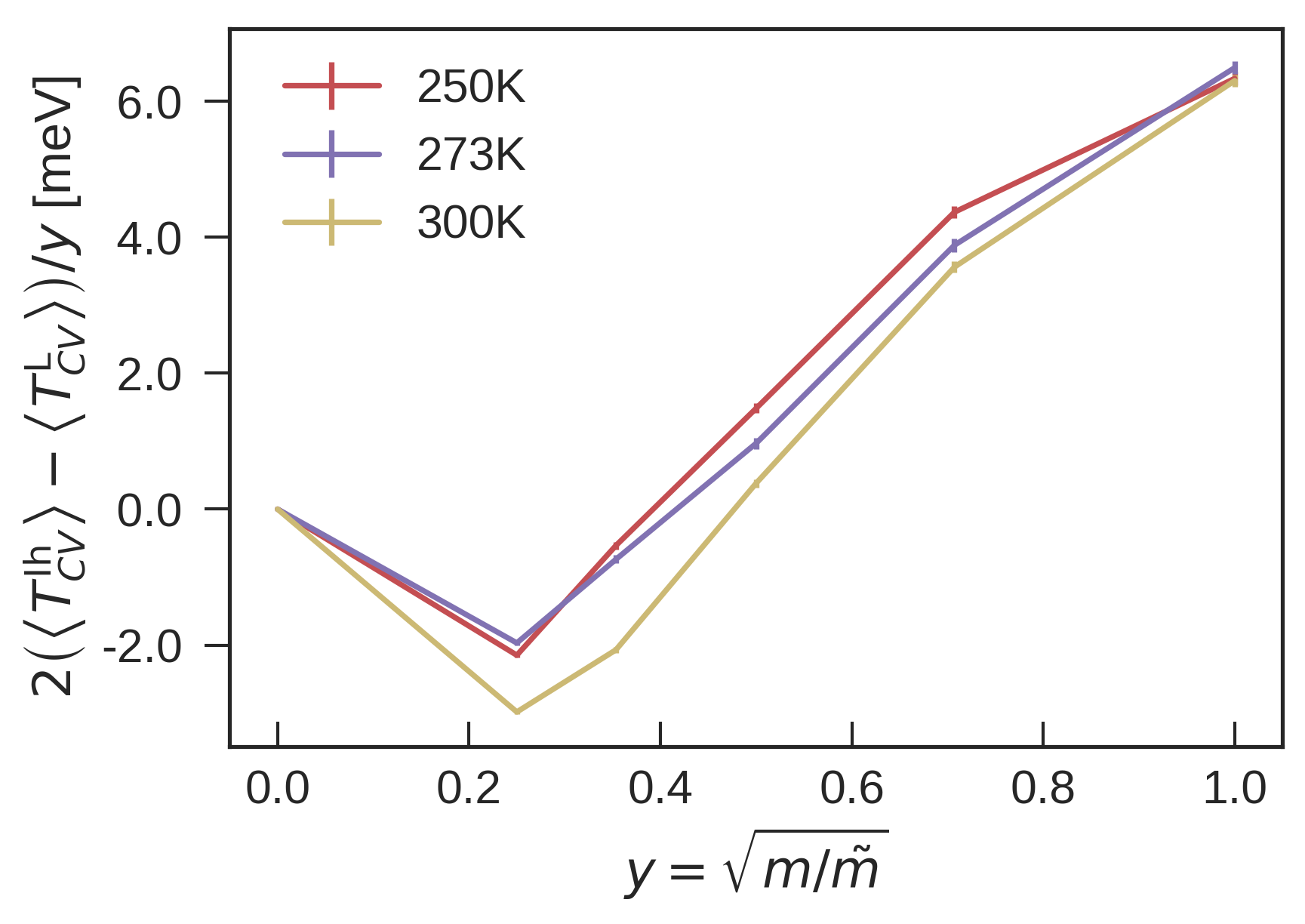}
\caption{
The integral from the classical limit to the full quantum treatment (~\eqref{eq:ti-u}), for the case of ice Ic and Ih (upper panel),
and ice Ih and liquid water (lower panel).
}
\label{fig:si2}
\end{figure}

\bc{NQEs lower the melting point of H$_2$O by about 8\,K compared with classical water.
The difference in $\tm$ between the D$_2$O and H$_2$O is predicted to be $8\pm2$~K,
consistent with the result obtained using the q-TIP4P/F water model~\cite{ramirez2010quantum},
and in rough agreement with experiment ($3.82$\,K)~\cite{bridgman1935pressure}.
Curiously, the $\tm$ of D$_2$O is about the same as the classical water.
To elucidate the reason,
we plot the the integral when performing TI from physical masses ($m_\mathrm{H}$ for H) to classical masses ($\infty$) in Fig.~\ref{fig:si2}.
It can be seen that NQEs initially,  from $m_\mathrm{H}$ to about $6m_\mathrm{H}$, stabilize water relative to ice.
Then, from $6m_\mathrm{H}$ to $\infty$ they stabilize ice. 
When performing TI from the atomic mass of deuterium to the classical mass, NQEs thus largely cancel out.
This reversal of NQEs at different atomic masses has been observed before for q-TIP4P/F water~\cite{ramirez2010quantum} and for stacked polyglutamine~\cite{rossi2015stability}, and has been interpreted as a manifestation of competing quantum effects.
}

\begin{table}[htb]
\centering
\caption{\bc{Predictions of the melting point ($\tm$) and the heat of fusion ($H_f$). The number in the bracket indicates the statistical uncertainty in the last digit.}}
\label{tab:gm}
\begin{tabular}{lrr}
\hline\hline
model &  $\tm$ [K] & $H_f$ [$\meVmolecule$] \\
\hline
 NN-classical & 279.6(4) & 67.8(2)\\
 DFT-classical & 275(2) & 58(2)\\
 DFT-H$_2$O & 267(2) & 52(3)\\
 DFT-D$_2$O & 275(2) & 58(2) \\
 experiment-H$_2$O & 273.15 & 62.3\\
 experiment-D$_2$O & 276.97 & 64.5\\
\hline\hline
\end{tabular}
\end{table}

\subsection*{Conclusions}

\bc{
We show that a revPBE0-D3 description of the electronic structure predicts properties for ice Ih, ice Ic and liquid water that are in excellent quantitative agreement with experiment.
This is made possible by
using a ML potential as an intermediate surrogate model,
and by employing advanced free energy techniques.
We not only rigorously compute but also quantitatively analyze the individual contributions from NQEs, proton disorder, and anharmonicity,

This study demonstrates that it is possible to achieve a sub-meV level of statistical accuracy in predicting the thermodynamic properties of a complex system such as water at a hybrid DFT level of theory.
The idea of using ML potentials as sampling devices significantly broadens the applicability and prowess of electronic structure
approaches,
making 
it affordable to utilize them in the accurate computations of free energies and other thermodynamic properties.
The overall framework and the free energy methods described here provide a general, accurate and robust way for first-principles predictions of thermodynamic properties of a plethora of physical systems,
such as pharmaceutical compounds, hydrogen storage materials, hydrocarbons, and metallic alloys.
}
\section*{Methods}
\subsection*{Simulation details}

The density isobar in
Fig.~\ref{fig:nn-density} is computed using both classical MD and PIMD simulations in the NPT ensemble for ice Ic, ice Ih, and liquid water systems of 64 molecules.
We have confirmed that the equilibrium density computed with 64 water molecules in classical molecular dynamics simulations is consistent with the values obtained for systems with about 2,000 molecules. 
All MD simulations and PIMD simulations that use 56 beads are performed employing the i-PI code~\cite{ipi2018} in conjunction with LAMMPS~\cite{plimpton1995lammps} with a NN potential implementation~\cite{Singraber2018nnp}.

\subsection*{Interface pinning}
The interface pinning simulations~\cite{pedersen2015computing} were performed using the PLUMED code~\cite{tribello2014plumed} 
on an ice-liquid system containing 5760 molecules at temperatures ranging from 250\,K to 300\,K and pressure 1\,bar,
employing the NN potential.

\subsection*{Accounting for NQEs}

NQEs on the chemical potential difference between ice Ic and ice Ih at $200$\,K, $250$\,K, $273$\,K and $300$\,K are taken into account by integrating the quantum centroid virial kinetic energy $\avg{T_{CV}}$ with respect to the fictitious  ``atomic'' mass $\ficm$ from the classical mass (i.e. infinity) to the physical masses of oxygen and hydrogen atoms
~\cite{ceriotti2013efficient,cheng2016nuclear,cheng2014direct,cheng2018hydrogen}.
, i.e.
\begin{equation}
        \micihnn-\micihclnn = \int_{m}^{\infty} d \ficm  \frac{\avg{T^{Ic}_{CV}(\ficm)}-\avg{T^{Ih}_{CV}(\ficm)}}{\ficm} 
        \label{eq:ti-u}
\end{equation}
where $m$ are the physical masses of the elements. 
In practice, a change of variable $y=\sqrt{m/\ficm}$ is applied to reduce the discretisation error in the evaluation of the integral~\cite{ceriotti2013efficient}, and the integrand is evaluated using PIMD simulations for $y=1/4, 1/2\sqrt{2}, 1/2, 1/\sqrt{2}, 1$, i.e.
\begin{equation}
\micihnn-\micihclnn =
2\int_0^{1} \frac{\avg{T^{Ic}_{CV}(1/y^2)}-\avg{T^{Ih}_{CV}(1/y^2)}}{y} d y.
\label{eq:ti-y}
\end{equation}
To evaluate this integral, we perform PIMD simulation hat use 56 beads at the NPT ensemble for systems containing 64 molecules.
For the case of ice Ih and liquid water, the treatment is similar.

\begin{figure}
\centering
\includegraphics[width=0.45\textwidth]{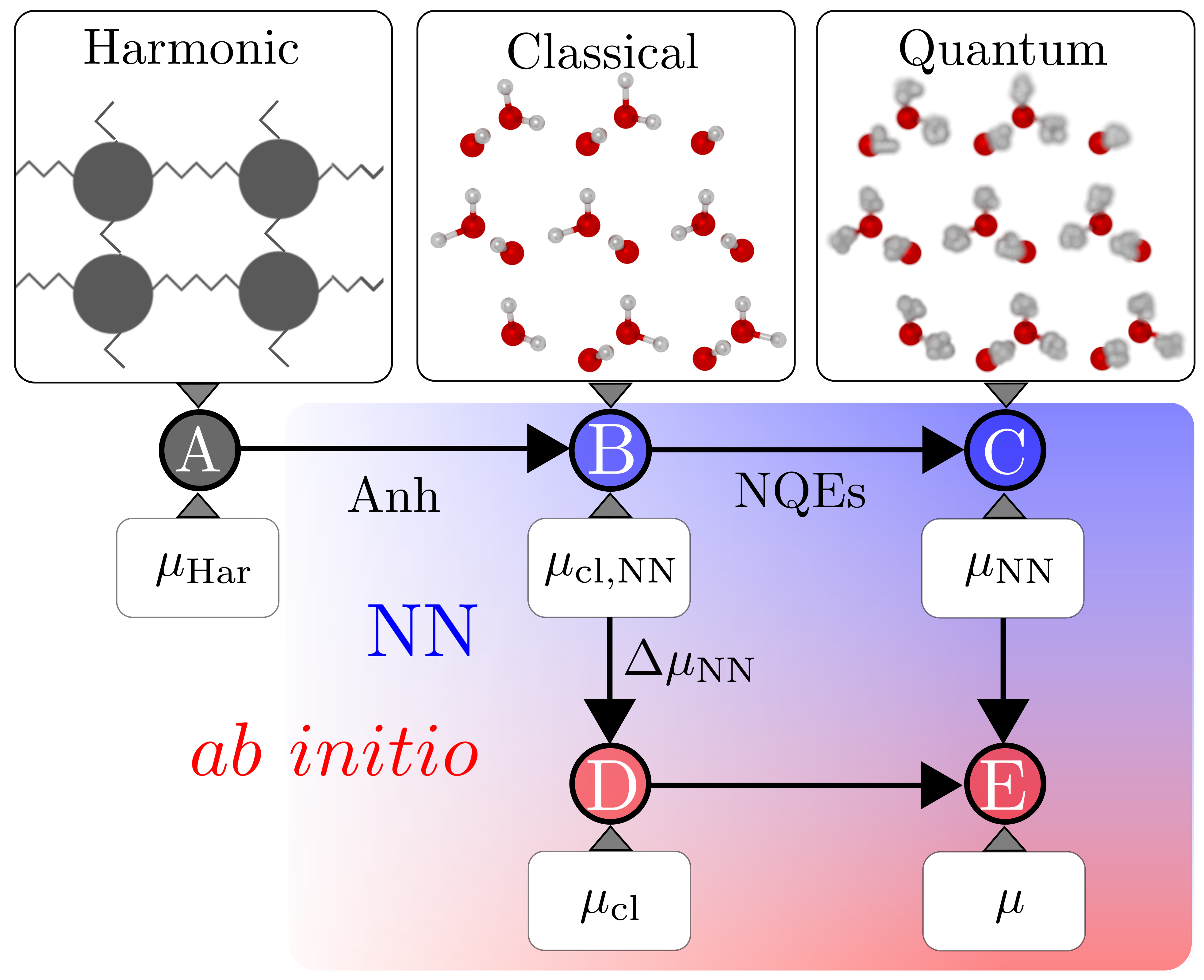}
\caption{
\bc{A schematic of the thermodynamic integration workflow, which starts from a harmonic reference crystal, uses a NN description as an intermediate step, and finally arrives at the underlying \emph{ab initio} level. 
The chemical potentials indicated here are related to the absolute Gibbs free energy of the systems by $\mu \equiv G/N$.}
}
\label{fig:ti_scheme}
\end{figure}

\subsection*{Workflow for computing $\micih$}
\bc{
Here we describe the workflow for computing absolute Gibbs free energy and thereby the chemical potential of an ice system.
The first step is a TI from a harmonic reference to a classical ice system ($A\to B$ in Fig.~\ref{fig:ti_scheme}).
We closely follow the methods described in Ref.~\cite{cheng2018computing}:
first integrate from a Debye crystal to classical ice at 25~K in the NVT ensemble, then transition to the NPT ensemble, and finally evaluate the temperature dependence of the Gibbs free energy using MD simulations in the NPT ensemble at temperatures between $25$\,K and $300$\,K.
Subsequently, to reach the \emph{ab initio} description of classical ice from the NN description ($B\to D$ in Fig.~\ref{fig:ti_scheme}),
the correction term $\Delta \mu_{\text{NN}}$ (see Fig.~\ref{fig:nn-dft-dif}) computed using the free energy perturbation expression in~\eqref{eq:fep}
is included.
Finally, to describe \emph{ab initio} ice with quantum-mechanical nuclei ($D\to E$ in Fig.~\ref{fig:ti_scheme}), 
NQEs are included by integrating from the infinite atomic mass to the physical masses (see \eqref{eq:ti-u}).
As an alternative strategy, one can also follow the TI route $A\to B \to C \to E$, 
but this requires re-weighting the whole ring-polymer system in PIMD simulations using ~\eqref{eq:fep}, which is more costly.
}

\section*{Acknowledgements}
We thank Matti Hellstr\"om for providing the VASP data and
Gerit Brandenburg for benchmarking the revPBE0+D3 functional.
BC would like to acknowledge funding from the Swiss National Science 
Foundation (Project ID 200021-159896),
 a partial funding of a visit in G\"ottingen by the International Center for Advanced Studies of Energy Conversion (ICASEC),
and generous allocation of CPU time by
CSCS under Project ID s787.
MC and EAE acknowledge funding by the European Research Council under the European Union's Horizon 2020 research and innovation programme (grant agreement no. 677013-HBMAP).
JB thanks the DFG for a Heisenberg Professorship (BE3264/11-2). CD acknowledges funding from the Austrian Science Fund (FWF) [SFB ViCoM, Project No. F41].


\begin{thebibliography}{10}

\bibitem{bartels2012ice}
Bartels-Rausch T, et~al. (2012) {Ice structures, patterns, and processes: A
  view across the ice-fields.}
\newblock {\em Review of Modern Physics} 84:885--944.

\bibitem{rall1985ice}
Rall WF, Fahy GM (1985) Ice-free cryopreservation of mouse embryos at- 196 c by
  vitrification.
\newblock {\em Nature} 313(6003):573.

\bibitem{murray2005formation}
Murray BJ, Knopf DA, Bertram AK (2005) {The formation of cubic ice under
  conditions relevant to the Earth's atmosphere.}
\newblock {\em Nature} 434:202--205.

\bibitem{lupi2017role}
Lupi L, et~al. (2017) Role of stacking disorder in ice nucleation.
\newblock {\em Nature} 551(7679):218.

\bibitem{kuhs2012extent}
Kuhs WF, Sippel C, Falenty A, Hansen TC (2012) {Extent and relevance of
  stacking disorder in ``ice Ic''.}
\newblock {\em Proceedings of the National Academy of Sciences}
  109:21259--21264.

\bibitem{vega2009ice}
Vega C, Abascal JL, Conde M, Aragones J (2009) What ice can teach us about
  water interactions: a critical comparison of the performance of different
  water models.
\newblock {\em Faraday discussions} 141:251--276.

\bibitem{morales2014qmc}
Morales MA, et~al. (2014) {Quantum Monte Carlo Benchmark of
  Exchange-Correlation Functionals for Bulk Water}.
\newblock {\em Journal of Chemical Theory and Computation} 10:2355.

\bibitem{zhang2011structural}
Zhang, C.;~Wu JGGGF (2011) {Structural and Vibrational Properties of Liquid
  Water from Van Der Waals Density Functionals}.
\newblock {\em Journal of Chemical Theory and Computation.} 7:3054.

\bibitem{santra2009coupledcluster}
Santra B, Michaelides A, Scheffler M (2009) {Coupled Cluster Benchmarks of
  Water Monomers and Dimers Extracted from Density-Functional Theory Liquid
  Water: The Importance of Monomer Deformations}.
\newblock {\em Journal of Chemical Physics} 131:124509.

\bibitem{ceriotti2018NQE}
Markland TE, Ceriotti M (2018) {Nuclear quantum effects enter the mainstream}.
\newblock {\em Nature Reviews Chemistry} 2:0109.

\bibitem{ramirez2014configurational}
Herrero CP, Ram\'{i}rez R (2014) {Configurational entropy of
  hydrogen-disordered ice polymorphs}.
\newblock {\em Journal of Chemical Physics} 140:234502.

\bibitem{raza2011proton}
Raza Z, et~al. (2011) {Proton ordering in cubic ice and hexagonal ice; a
  potential new ice phase -- XIc.}
\newblock {\em Physical Chemistry Chemical Physics} 13:19788--19795.

\bibitem{engel2015anharmonic}
Engel EA, Monserrat B, Needs RJ (2015) Anharmonic nuclear motion and the
  relative stability of hexagonal and cubic ice.
\newblock {\em Physical Review X} 5(2):021033.

\bibitem{lekner1998energetics}
Lekner J (1998) {Energetics of hydrogen ordering in ice.}
\newblock {\em Physica B: Condensed Matter} 252:149--159.

\bibitem{ramirez2010quantum}
Ram\'{i}rez R, Herrero CP (2010) {Quantum path integral simulation of isotope
  effects in the melting temperature of ice Ih}.
\newblock {\em Journal of Chemical Physics} 133:144511.

\bibitem{habershon2009competing}
Habershon S, Markland TE, Manolopoulos DE (2009) {Competing quantum effects in
  the dynamics of a flexible water model.}
\newblock {\em Journal of Chemical Physics} 131:024501.

\bibitem{pamuk2012anomalous}
Pamuk B, et~al. (2012) {Anomalous Nuclear Quantum Effects in Ice.}
\newblock {\em Physical Review Letters} 108:193003.

\bibitem{cheng2016nuclear}
Cheng B, Behler J, Ceriotti M (2016) Nuclear quantum effects in water at the
  triple point: Using theory as a link between experiments.
\newblock {\em The journal of physical chemistry letters} 7(12):2210--2215.

\bibitem{ramirez2012qha}
Ram\'{i}rez R, Neuerburg N, Fern\'{a}ndez-Serra MV, Herrero CP (2012)
  {Quasi-harmonic approximation of thermodynamic properties of ice Ih, II, and
  III.}
\newblock {\em Journal of Chemical Physics} 137:044502.

\bibitem{engel2018firstprinciples}
Engel EA, Li Y, Needs RJ (2018) {First-principles momentum distributions and
  vibrationally corrected permittivities of hexagonal and cubic ice}.
\newblock {\em Physical Review B} 97:054312.

\bibitem{zhang1998gga}
Zhang Y, Yang W (1998) {Comment on ``Generalized Gradient Approximation Made
  Simple''}.
\newblock {\em Physical Review Letters} 80:890.

\bibitem{adamo1999pbe0}
Adamo C, Barone V (1999) {Toward reliable density functional methods without
  adjustable parameters: The PBE0 model}.
\newblock {\em Journal of Chemical Physics} 110:6158.

\bibitem{goerigk2011revpb0d3}
Goerigk L, Grimme S (2011) {A thorough benchmark of density functional methods
  for general main group thermochemistry, kinetics, and noncovalent
  interactions}.
\newblock {\em Physical Chemistry Chemical Physics} 13:6670.

\bibitem{grimme2010d3}
Grimme S, Antony J, Ehrlich S, Krieg S (2010) {A consistent and accurate ab
  initio parametrization of density functional dispersion correction (dft-d)
  for the 94 elements H-Pu}.
\newblock {\em Journal of Chemical Physics} 132:154104.

\bibitem{goerigk2011thorough}
Goerigk L, Grimme S (2011) A thorough benchmark of density functional methods
  for general main group thermochemistry, kinetics, and noncovalent
  interactions.
\newblock {\em Physical Chemistry Chemical Physics} 13(14):6670--6688.

\bibitem{marsalek2017dynamics}
Marsalek O, Markland TE (2017) {Quantum Dynamics and Spectroscopy of Ab Initio
  Liquid Water: The Interplay of Nuclear and Electronic Quantum Effects}.
\newblock {\em J. Phys. Chem. Lett.} 8:1545.

\bibitem{macher2014random}
Macher M, Klime{\v{s}} J, Franchini C, Kresse G (2014) The random phase
  approximation applied to ice.
\newblock {\em The Journal of Chemical Physics} 140(8):084502.

\bibitem{grabowski2009ab}
Grabowski B, Ismer L, Hickel T, Neugebauer J (2009) Ab initio up to the melting
  point: Anharmonicity and vacancies in aluminum.
\newblock {\em Physical Review B} 79(13):134106.

\bibitem{glensk2014breakdown}
Glensk A, Grabowski B, Hickel T, Neugebauer J (2014) Breakdown of the arrhenius
  law in describing vacancy formation energies: the importance of local
  anharmonicity revealed by ab initio thermodynamics.
\newblock {\em Physical Review X} 4(1):011018.

\bibitem{behler2007generalized}
Behler J, Parrinello M (2007) Generalized neural-network representation of
  high-dimensional potential-energy surfaces.
\newblock {\em Physical Review Letters} 98(14):146401.

\bibitem{behler2017anie}
Behler J (2017) First principles neural network potentials for reactive
  simulations of large molecular and condensed systems.
\newblock {\em Angew. Chem. Int. Ed.} 56(42):12828.

\bibitem{morawietz2016van}
Morawietz T, Singraber A, Dellago C, Behler J (2016) How van der waals
  interactions determine the unique properties of water.
\newblock {\em Proceedings of the National Academy of Sciences}
  113(30):8368--8373.

\bibitem{runner}
Behler J (2018) {\em RuNNer -- A Neural Network Code for High-Dimensional
  Neural Network Potentials}.
\newblock (Universit\"at G\"ottingen).

\bibitem{lippert1999cp2k}
Lippert G, Hutter J, Parrinello M (1999) {The Gaussian and augmented-plane-wave
  density functional method for ab initio molecular dynamics simulations}.
\newblock {\em Theoretical Chemistry Accounts} 103:124.

\bibitem{kresse1996software}
Kresse G (1996) Software vasp, vienna, 1999; g. kresse, j. furthm{\"u}ller.
\newblock {\em Phys. Rev. B} 54(11):169.

\bibitem{hare1987density}
Hare D, Sorensen C (1987) The density of supercooled water. ii. bulk samples
  cooled to the homogeneous nucleation limit.
\newblock {\em The Journal of chemical physics} 87(8):4840--4845.

\bibitem{skinner2014structure}
Skinner LB, Benmore C, Neuefeind JC, Parise JB (2014) The structure of water
  around the compressibility minimum.
\newblock {\em The Journal of chemical physics} 141(21):214507.

\bibitem{soper2000radial}
Soper A (2000) The radial distribution functions of water and ice from 220 to
  673 k and at pressures up to 400 mpa.
\newblock {\em Chemical Physics} 258(2-3):121--137.

\bibitem{chen2016ab}
Chen W, Ambrosio F, Miceli G, Pasquarello A (2016) Ab initio electronic
  structure of liquid water.
\newblock {\em Physical review letters} 117(18):186401.

\bibitem{bridgman1935pressure}
Bridgman P (1935) The pressure-volume-temperature relations of the liquid, and
  the phase diagram of heavy water.
\newblock {\em The Journal of Chemical Physics} 3(10):597--605.

\bibitem{roettger1994lattice}
R{\"{o}}ttger K, Endriss A, Ihringer J, Doyle S, Kuhs WF (1994) {Lattice
  constants and thermal expansion of H$_2$O and D$_2$O ice Ih between 10 and
  265 K}.
\newblock {\em Acta Crystallographica B} 50:644.

\bibitem{morrone2008nuclear}
Morrone JA, Car R (2008) Nuclear quantum effects in water.
\newblock {\em Physical review letters} 101(1):017801.

\bibitem{mantz2006structural}
Mantz YA, Chen B, Martyna GJ (2006) Structural correlations and motifs in
  liquid water at selected temperatures: Ab initio and empirical model
  predictions.
\newblock {\em The Journal of Physical Chemistry B} 110(8):3540--3554.

\bibitem{ceriotti2011accelerating}
Ceriotti M, Manolopoulos DE, Parrinello M (2011) Accelerating the convergence
  of path integral dynamics with a generalized langevin equation.
\newblock {\em The Journal of chemical physics} 134(8):084104.

\bibitem{ceriotti2016nqe}
Ceriotti M, et~al. (2016) {Nuclear Quantum Effects in Water and Aqueous
  Systems: Experiment, Theory, and Current Challenges}.
\newblock {\em Chemical Reviews} 116:7529.

\bibitem{spura2015nuclear}
Spura T, John C, Habershon S, K{\"u}hne TD (2015) Nuclear quantum effects in
  liquid water from path-integral simulations using an ab initio force-matching
  approach.
\newblock {\em Molecular Physics} 113(8):808--822.

\bibitem{cisneros2016modeling}
Cisneros GA, et~al. (2016) Modeling molecular interactions in water: from
  pairwise to many-body potential energy functions.
\newblock {\em Chemical reviews} 116(13):7501--7528.

\bibitem{behler2015constructing}
Behler J (2015) Constructing high-dimensional neural network potentials: A
  tutorial review.
\newblock {\em International Journal of Quantum Chemistry} 115(16):1032--1050.

\bibitem{matsumoto2018genice}
Matsumoto M, Yagasaki T, Tanaka H (2018) Genice: Hydrogen-disordered ice
  generator.
\newblock {\em Journal of computational chemistry} 39(1):61--64.

\bibitem{grau2014qmsimulation}
Drechsel-Grau C, Marx D (2014) {Quantum Simulation of Collective Proton
  Tunneling in Hexagonal Ice Crystals}.
\newblock {\em Physical Review Letters} 112:148302.

\bibitem{cheng2018computing}
Cheng B, Ceriotti M (2018) {Computing the absolute Gibbs free energy in
  atomistic simulations: Applications to defects in solids}.
\newblock {\em Physical Review B} 97:054102.

\bibitem{molinero2008water}
Molinero V, Moore EB (2008) Water modeled as an intermediate element between
  carbon and silicon.
\newblock {\em The Journal of Physical Chemistry B} 113(13):4008--4016.

\bibitem{cheng2018theoretical}
Cheng B, Dellago C, Ceriotti M (2018) Theoretical prediction of the homogeneous
  ice nucleation rate: disentangling thermodynamics and kinetics.
\newblock {\em arXiv preprint arXiv:1807.05551}.

\bibitem{reddy2016accuracy}
Reddy SK, et~al. (2016) On the accuracy of the mb-pol many-body potential for
  water: Interaction energies, vibrational frequencies, and classical
  thermodynamic and dynamical properties from clusters to liquid water and ice.
\newblock {\em The Journal of chemical physics} 145(19):194504.

\bibitem{carr2014spectroscopic}
Carr THG, Shephard JJ, Salzmann CG (2014) {Spectroscopic Signature of Stacking
  Disorder in Ice I.}
\newblock {\em Journal of Physical Chemistry Letters} 5:2469--2473.

\bibitem{pedersen2015computing}
Pedersen UR, Hummel F, Dellago C (2015) Computing the crystal growth rate by
  the interface pinning method.
\newblock {\em The Journal of Chemical Physics} 142(4):044104.

\bibitem{haji2015direct}
Haji-Akbari A, Debenedetti PG (2015) Direct calculation of ice homogeneous
  nucleation rate for a molecular model of water.
\newblock {\em Proceedings of the National Academy of Sciences}
  112(34):10582--10588.

\bibitem{espinosa2014homogeneous}
Espinosa J, Sanz E, Valeriani C, Vega C (2014) Homogeneous ice nucleation
  evaluated for several water models.
\newblock {\em The Journal of Chemical Physics} 141(18):18C529.

\bibitem{rossi2015stability}
Rossi M, Fang W, Michaelides A (2015) Stability of complex biomolecular
  structures: van der waals, hydrogen bond cooperativity, and nuclear quantum
  effects.
\newblock {\em The journal of physical chemistry letters} 6(21):4233--4238.

\bibitem{ipi2018}
Kapil V, et~al. (2018) i-pi 2.0: A universal force engine for advanced
  molecular simulations.
\newblock {\em Computer Physics Communications}.

\bibitem{plimpton1995lammps}
Plimpton S (1995) {Fast Parallel Algorithms for Short-Range Molecular
  Dynamics}.
\newblock {\em Journal of Computational Physics} 117:1.

\bibitem{Singraber2018nnp}
Singraber A, Behler J, Dellago C (2018) A library-based lammps implementation
  of high-dimensional neural network potentials.
\newblock {\em submitted to the Journal of Chemical Theory and Computation}.

\bibitem{tribello2014plumed}
Tribello GA, Bonomi M, Branduardi D, Camilloni C, Bussi G (2014) Plumed 2: New
  feathers for an old bird.
\newblock {\em Computer Physics Communications} 185(2):604--613.

\bibitem{ceriotti2013efficient}
Ceriotti M, Markland TE (2013) Efficient methods and practical guidelines for
  simulating isotope effects.
\newblock {\em The Journal of chemical physics} 138(1):014112.

\bibitem{cheng2014direct}
Cheng B, Ceriotti M (2014) Direct path integral estimators for isotope
  fractionation ratios.
\newblock {\em The Journal of chemical physics} 141(24):244112.

\bibitem{cheng2018hydrogen}
Cheng B, Paxton AT, Ceriotti M (2018) Hydrogen diffusion and trapping in
  $\alpha$-iron: The role of quantum and anharmonic fluctuations.
\newblock {\em Physical review letters} 120(22):225901.

\end{thebibliography}
\end{document}